\documentclass[12pt,preprint]{aastex}

\def\etal{{\it et al.} }

\begin{document}
\title{A two component jet model for the X-ray afterglow flat segment
in short GRB 051221A}

\author{Z.P.Jin$^{1,2}$, T. Yan$^{1,2}$, Y. Z. Fan$^{1,2,3,4}$ and D. M.
Wei$^{1,2,5}$} \affil{$^1$ Purple Mountain Observatory, Chinese
Academy
of Science, Nanjing 210008, China.\\
$^2$ National Astronomical Observatories,Chinese Academy of
Sciences, Beijing, 100012, China.\\
$^3$ The Racah Inst. of Physics, Hebrew University, Jerusalem
91904, Israel.\\
$^4$ Golda Meir Fellow.\\
$^5$ Joint Center for Particle Nuclear Physics and Cosmology of
Purple Mountain Observatory - Nanjing University, Nanjing
210008,China.}

\begin{abstract}
In the double neutron star merger or neutron star-black hole merger
model for short GRBs, the outflow launched might be mildly
magnetized and neutron rich. The magnetized neutron-rich outflow
will be accelerated by the magnetic and thermal pressure and may
form a two component jet finally, as suggested by Vlahakis, Peng \&
K\"{o}nigl (2003). We show in this work that such a two component
jet model could well reproduce the multi-wavelength afterglow
lightcurves, in particular the X-ray flat segment, of short GRB
051221A. In this model, the central engine need not to be active
much longer than the prompt $\gamma-$ray emission.
\end{abstract}


\section{Introduction}
\label{sect:Intro}

Kouveliotou et al. (1993) showed that the Gamma-ray Bursts (GRBs)
can be divided into two distinguished groups: long GRBs (LGRBs) with
duration $\geq$ 2 sec and short GRBs (SGRBs) with duration $<$ 2
sec. Fruitful afterglow data have been collected for LGRBs and
support the collapsar model (see Woosley \& Bloom 2006 for a recent
review). The afterglow data of SGRBs are relatively rare and have
not been reliably detected before May 2005 \cite{Gehrels05}. The
leading model for SGRBs is the merger of neutron stars or neutron
star--black hole binaries \cite{Eichler89, Narayan92}. One
attractive feature of such a compact star binaries merger model is
that the accretion disk formed is too small to give rise to a long
burst, provided that the central remnant has collapsed to a black
hole immediately (e.g., Narayan \etal 2001: Rosswog et al. 2002;
Aloy \etal 2005; cf. Klu\'zniak \& Ruderman 1998; Rosswog \etal
2003). Recently, this view of SGRB origin is supported by
observations (see M\'{e}sz\'{a}ros 2006 for a recent review): a) For
some short bursts, the host is early type galaxy, with a stellar
population older than $\sim 1$ Gyr. The overall star formation rate
is very low \cite{Fox05, Barthelmy05}. b) Deep search of an
underlying supernova has resulted in null results (e.g. Hjorth \etal
2005). c) The typical number density of the medium surrounding the
SGRB progenitors is low to $10^{-3}~{\rm cm^{-3}}$ and a stellar
wind profile is disfavored.

While the compact objects binary merger model for SGRBs is largely
consistent with the ongoing afterglow observations, some lightcurve
data, for example, the optical or X-ray flares in GRB050709 (Fox
\etal 2005) and GRB050724 \cite{Barthelmy05}, and the X-ray
afterglow flat segment in GRB051221A \cite{Soderberg06, Burrows06},
have not been well interpreted. One speculation is that the central
engine still works after the cease of the prompt $\gamma-$ray
emission \cite{Fan05, Zhang06, Dai06, Gao06, Fan06b} and may involve
some magnetic processes. For example, Fan \& Xu (2006) suggested a
magnetar wind energy injection model for the X-ray afterglow flat
segment lasting $\sim 10^4$ seconds in GRB 051221A. However, it is
still unclear that whether the supermassive or hypermassive magnetar
formed in the double neutron star coalescence (Price \& Rosswog
2006) could survive for such a long time \cite{Duez06}. In this
Letter we suggest a two component jet model to fit the
multi-wavelength afterglow data of GRB 051221A. In this model, the
central engine need not to be active much longer than the prompt
$\gamma-$ray emission. We discuss the possible physical scenario
giving rise to such a two component jet in section 2. We fit to the
afterglow data numerically in section 3. We summarize our result
with some discussions in section 4.

\section{A two component jet model}
\label{sect:model}

In the double neutron stars or neutron star--black hole merger
scenario, the outflow is very likely to be neutron rich and might
also be magnetized \cite{Price06}. For a magnetized neutron-rich
outflow, the acceleration process is significantly different from
that of the standard fireball \cite{Piran93}. As shown in Vlahakis,
Peng \& K\"{o}nigl (2003), the neutron-to-proton ratio in disk-fed
outflows might be as high as $\sim 30$ and the neutrons can decouple
at a Lorentz factor $\Gamma \sim$ tens, while the protons  continue
to be accelerated to a $\Gamma \sim$ hundreds and is collimated to a
narrower angle by the electromagnetic force. As a result, a two
component jet might be formed\cite{Vlahakis03, Peng05}. We use such
a structured jet profile to fit the afterglow of short GRB051221A in
next section.

\section{Fit to the afterglow light curve of GRB 051221A}
\label{sect:lightcurves}

\subsection{The observation}
GRB051221A located by {\it Swift} is one of the few SGRBs whose
afterglows were well detected in multi-bands. Its duration is
$T_{90}=1.4\pm0.2$s \cite{Cummings05}, the peak energy is $E_{\rm
p}=402_{-72}^{+93}$keV \cite{Burrows05}, the gamma-ray (20keV-2MeV)
fluence is about $3.2\times10^{-6}{\rm erg\cdot cm}^{-2}$
\cite{Golenetskii05}. With the optical spectrometric measured
redshift $z=0.5464$ \cite{Soderberg06}, the total isotropic
gamma-ray (20keV-2MeV) energy of GRB051221A is about
$2.4\times10^{51}$ erg, make it the most fluent SGRB and about
$6-35$ times higher than the others \cite{Soderberg06}. The Swift
XRT began observing GRB051221A 88 seconds after the burst and found
the bright, rapidly fading X-ray afterglow \cite{Burrows05}. The
X-ray afterglow early decay index is $\alpha_{\rm
X1}=1.16_{-0.17}^{+0.09}$ (100-2200s) and late $\alpha_{\rm
X3}=1.09_{-0.09}^{+0.10}$ ($3\times10^{4}-3.6\times10^{5}$s),
between them is a slow decay decline $\alpha_{\rm
X2}=0.04_{-0.21}^{+0.27}$, and the average photon spectral index of
these three segments is $\Gamma_{\rm X}=2.1\pm0.2$(Burrows \etal
2006). The Chandra observations show a steep decay $\alpha_{\rm
X4}=1.93_{-0.19}^{+0.23}$ after $\sim3.6\times10^{5}$s, and the
average photon spectral index of this segment is $\Gamma_{\rm
X}=1.94_{-0.19}^{+0.29}$ \cite{Burrows06}. The GMOS start optical
observation in $r'$ band 2.8 hours after burst and get an optical
decline $\alpha_{\rm opt}=0.92\pm0.04$ \cite{Soderberg06}.

The spectral analysis of the optical-to-X-ray afterglow data at
$\sim 1$ day after the trigger of the burst suggests a cooling
frequency $\nu_{\rm c} \sim 2\pm1\times10^{17}$ Hz
\cite{Soderberg06}. With the early X-ray decay index (100-2200s),
we have an electron index $p=1-4\alpha_{\rm
X1}/3=2.55_{-0.23}^{+0.12}$, note that $\nu_{\rm c}\propto
t^{-1/2}$ is over the XRT band. And
later($3\times10^{4}-3.6\times10^{5}$s), when $\nu_{\rm c}$ is
crossing the X-ray band, the decay index smoothly changes from
$(3-3p)/4$ to $(2-3p)/4$. The average decay index $\alpha_{\rm
X3}$ is between these two values, that we have
$2.12_{-0.12}^{+0.13}=(2-4\alpha_{\rm X3})/3\leq
p\leq1-4\alpha_{\rm X3}/3=2.45_{-0.12}^{+0.13}$. Or by the optical
decay index at the same time, we have $p=1-4\alpha_{\rm
opt}/3=2.23\pm0.05$ ($\nu_{\rm c}$ is still above the optical band
now). It is interesting to note that the early and later electron
distribution indexes $p$ seem to be much different (see also
Burrows \etal 2006 for a similar argument). In previous numerical
fit to the afterglow data of GRB 051221A, a constant $p$ can not
well reproduce both the early and the late X-ray declines, as
shown in Figure 3 of Burrows \etal (2006) and Figure 1 of Fan \&
Xu (2006). In our two component jet model, the shock parameters
($\epsilon_e,~\epsilon_B,~p$) might be different for the narrow
and the wide components. The early afterglow emission is mainly
contributed by the narrow jet component while the late afterglow
is dominated by the forward shock emission of the wide jet
component. So the values of $p$ for the early and later afterglow
data could be different.

\subsection{Numerical method}
We interpret the apparent flattening in the X-ray light curve being
caused by the emergence of the forward shock emission of the wide
jet component. The code here to fit the multi-band lightcurves has
been used in Yan \etal (2007), in which the outflow dynamical
evolution is governed by a set of differential equations.

{\bf Reverse shock phase.} In this case, the energy conservation of
the whole system that contains four regions (1) the unshocked
medium, (2) the shocked medium, (3) the shocked ejecta, and (4) the
unshocked ejecta (reads Huang et al. 2000;  Yan \etal 2007)

\begin{equation}
d(E_2+E_3+E_4)=-\varepsilon_2\gamma_2dU_2-\varepsilon_3\gamma_3dU_3,
\end{equation}
where $E_{\rm i}$ is kinetic energy measured in observer frame and
they are governed by
$E_2=(\gamma_2-1)m_2c^2+(1-\varepsilon_2)\gamma_2U_2$,
$E_3=(\gamma_3-1)m_3c^2+(1-\varepsilon_3)\gamma_3U_3$ and
$E_4=(\eta-1)(m_{\rm ej}-m_3)c^2$. Where $U_2=(\gamma_2-1)m_2c^2$
and $U_3=(\gamma_{34}-1)m_3c^2$ are internal energy measured in
the comoving frame. Note that in this subsection, the subscripts
$i=1-4$ represent the $i-$th region, respectively.
$\varepsilon_{\rm i}$ is the radiation fraction of internal
energy, $\beta_{\rm i}$, $\gamma_{\rm i}$ are the velocities (in
units of $c$) and the corresponding Lorentz factors measured in
observer frame, respectively; $m_{\rm i}$ is the rest mass of the
material, $\gamma_{34}$ is the Lorentz factor of region 3 relative
to region 4.

The mass of the medium swept by the forward shock is
\begin{equation}
dm_2=4\pi R^2n_1m_pdR,
\end{equation}
where $n_{\rm i}$ is the proton number density measured in the
comoving frame, $m_{\rm p}$ is the proton's rest mass and $R$ is
the radius of forward shock front to the central engine.

The mass of the medium swept by the reverse shock is
\begin{equation}
dm_3=4\pi R^2\eta \,n_4m_p(\beta_4-\beta_{\rm RS})dR,
\end{equation}
where $\eta$ is the initial Lorentz factor of outflow, and
$\beta_{\rm RS}\approx (\gamma_3 n_3 \beta_3-\gamma_4 n_4
\beta_4)/(\gamma_3 n_3-\gamma_4 n_4)$ \cite{Sari95,Fan04} is the
velocity of reverse shock.

With eq.(1-3) and assuming that $\gamma_2=\gamma_3$, we have the
following equation to describe the dynamical evolution of forward
shock
\begin{equation}
d\gamma_2=-4\pi
R^2\frac{(\gamma_2^2-1)n_1m_p+(\gamma_2\gamma_{34}-\eta)\eta
n_4m_p(\beta_4-\beta_{\rm RS})}{I},
\end{equation} where $I=m_2+m_3+(1-\varepsilon_2)(2\gamma_2-1)m_2+(1-\varepsilon_3)
(\gamma_{34}-1)m_3+(1-\varepsilon_3)\gamma_2m_3
[\eta(1-\beta_2\beta_4)-\frac{\eta\beta_4}{\gamma_2^2\beta_2}]$.

{\bf After the cease of the reverse shock.} In this case, the
dynamical evolution of the forward shock has been discussed in
great detail \cite{Huang00}. The reverse shock electrons cool
adiabatically. We take the analytical approach of Sari \& Piran
(1999) to calculate the emission of those electrons.

In the afterglow lightcurve calculation (see Yan \etal 2007 for
details), the cooling of the shocked electrons due to both
synchrotron and inverse Compton has been taken into account
\cite{Sari98,Wei98,Sari01}. The synchrotron radiation of the forward
shock electrons on the ``equal arriving surface" (on which the
emission reaches us at the same time) has been integrated strictly
\cite{Huang00}.

\subsection{Fit to the afterglow data of GRB051221A}
\label{sec:Fit}

Using the code described above, we fit the afterglow of GRB051221a.
Here we consider two cases. First, we assume that the shock physical
parameters of two components are the same
\cite{Berger03,Sheth03,Huang04}. The parameters taken into account
are as follow: for the narrow component, the isotropic energy
$E_{\rm k,iso} =3.2\times10^{51}$ ergs, the initial Lorentz factor
$\eta=500$, and the half opening angle $\theta_j=0.03$; for the wide
component, $E_{\rm k,iso} =9.5\times10^{51}$ ergs, $\eta=50$, and
$\theta_j=0.1$. For both components, $\epsilon_e=0.3$,
$\epsilon_B=0.01$, and $p=2.2$. However, from Fig.1 we note that in
this case the fit is not very well, the reason is that we take the
same value of $p$ for the two components (similar conclusion can be
found in Burrows \etal 2006 and Fan \& Xu 2006). So we take a
different value of $p=2.8$ for the narrow component and find that
the numerical results can well reproduces the afterglow of
GRB051221A. In this case the ratio of total energy of narrow and
wide component is $1 : 30$, which is very different from $3:7$, the
ratio found in Vlahakis et al.'s simulation (2003). Second, we
assume that for the narrow and the wide components, the shock
parameters are different. This assumption is somewhat unusual. For
the pre-{\it Swift} GRBs, the best fitted micro physical parameters
are different from burst to burst \cite{Panaitescu01,Yost03}. For
the {\it Swift} GRBs, the situation is less clear because the
afterglows are usually too peculiar to be reasonably reproduced in
the standard fireball model (e.g., Fan \& Piran 2006). Figure 2
Shows our numerical fit to the X-ray and optical afterglows with the
following parameters: for the narrow component, the isotropic energy
$E_{\rm k,iso} =3.2\times10^{52}$ ergs, the initial Lorentz factor
$\eta=500$, $\theta_j=0.03$, $\epsilon_e=0.09$, $\epsilon_B=0.003$,
and $p=2.8$; for the wide component, $E_{\rm k,iso}
=9.5\times10^{51}$ ergs, $\eta=50$, $\theta_j=0.1$,
$\epsilon_e=0.3$, $\epsilon_B=0.01$, and $p=2.2$.

\section{Summary and Discussion}
\label{sect:summary}

In this work we show that the multi-band afterglows of GRB051221A
could be well fitted by a two component jet model. Such a two
component jet might be formed if the disc-fed outflow is neutron
rich and magnetized \cite{Vlahakis03}. Our good fit to the afterglow
data in turn suggests that the initial ejecta launched in the double
neutron stars merger or neutron star-black hole merger may be
neutron rich and magnetized. However, we point out that any model
consisting of a fast and a slower jet can also provide a good fit to
the observed data for 051221A. This and some other possibilities are
described below.

There are several possible ways to produce a slower jet: 1) Neutrino
annihilation and magnetic field were considered as the energy source
for SGRB in NS binary merger scenario \cite{Eichler89, Narayan92}.
Maybe they both operate, but only one mechanism produced the GRB
prompt emission while the other contribute only to late afterglow as
a second component \cite{Rosswog03, Price06}. 2)A thick disc acts as
high-density wall was considered to collimate GRB outflow
\cite{Rosswog03}. It was thought if mixing instabilities at the
walls spoil the jet with baryons, it would result in a
non-relativistic "dirty" fireball \cite{Rosswog03}. We consider that
if the disc changes to unstable when the GRB outflow launch, the
consequent outflow might be more baryon loaded and less collimated,
maybe as a second component. It should be noted that the validity of
these processes are quite uncertain and more detailed investigation
are needed before they can be used for afterglow calculations.

We would like to point out that the two component jet model is not
unique to account for the afterglow data of GRB051221A. One
possibility is that the GRB outflow might have a range of bulk
Lorentz factors and the inner parts moving with a bulk Lorentz
factor $\sim$ tens carry most of the energy \cite{Rees98}. As the
outer faster parts got decelerated by the external medium, the
inner parts would catch up with the decelerating forward shock and
increase its kinetic energy. For the particular GRB 051221A, an
energy--lorentz factor distribution $E(>\Gamma\sim 20)\propto
\Gamma^{-4.5}$ is needed to reproduce the X-ray afterglow flat
segment \cite{Soderberg06,Burrows06}.  But the physical process
pumping such kind of energy injection, in particular in the SGRB
scenario, is not clear yet. The other possibility is that the
central engine is active much longer than the prompt $\gamma-$ray
emission phase. For example, Fan \& Xu (2006) proposed a magnetar
wind energy injection model to account for the long term X-ray
flat segment.  However, it is still unclear that whether the
supermassive or hypermassive magnetar formed in the double neutron
star coalescence could survive for such a long time. In principle,
there could be a method to distinguish the two component jet model
from these two possibilities---In the two component jet model, the
first jet break might be observed when the edge of the narrow
component enters our line of sight. The lack of a detection of the
first jet break renders the energy injection model possibly. The
future fruitful muti-wavelength afterglow data may help us to pin
down the underlying physical processes.

\section*{Acknowledgments}

We thank the referee for useful comments which helped to improve
this paper. This work is supported by the National Natural Science
Foundation (grants 10225314, 10233010, 10621303 and 10673034) of
China, and the National 973 Project on Fundamental Researches of
China (NKBRSF G19990754).

\clearpage

\begin{figure}
\plotone{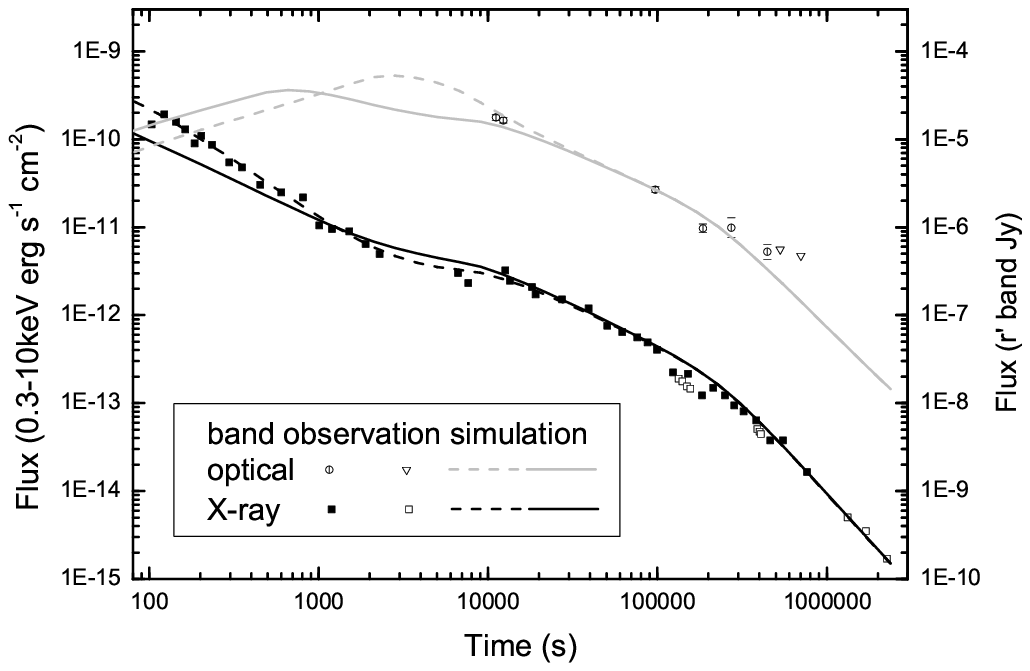}
\caption{Fit to the multi-band afterglow of
GRB051221A. Circles and squares are optical($r'$ band) and X-ray
(0.3-10keV) observations. The solid lines are our calculation. To
make a comparison, we also take a different $p=2.8$ for the narrow
jet, as the dashed lines shown.} \label{Fig:fig1}
\end{figure}

\clearpage

\begin{figure}
\plotone{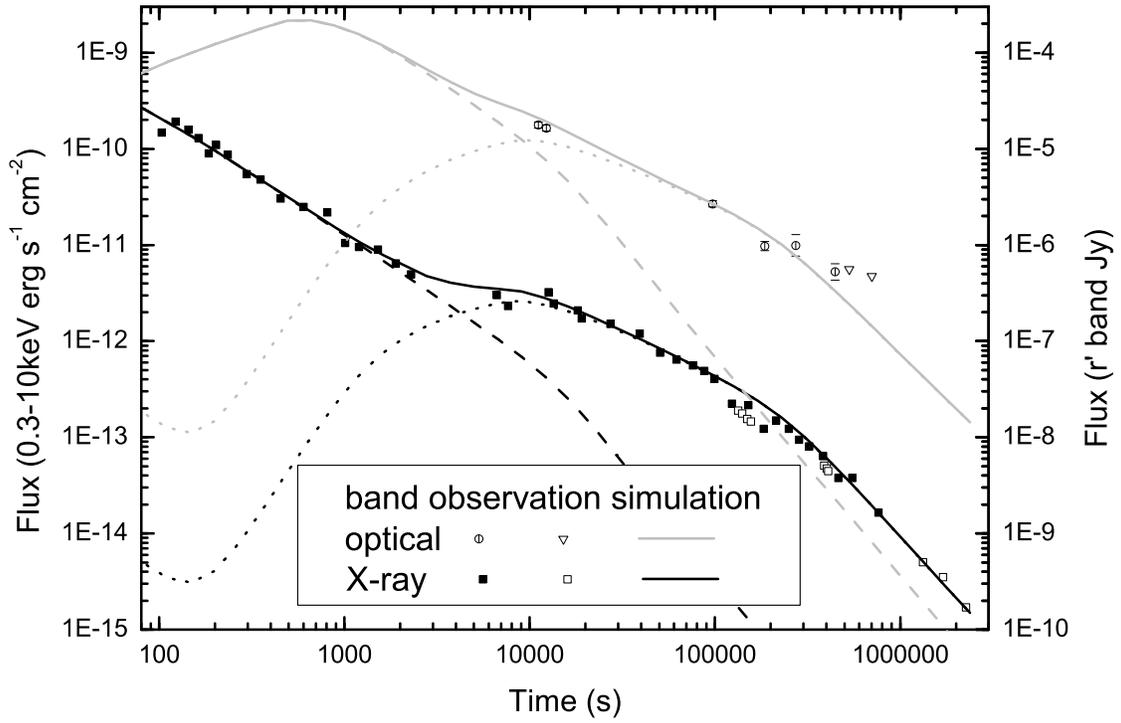}
\caption{Fit to the multi-band afterglow of
GRB051221A. Circles and squares are optical($r'$ band) and X-ray
(0.3-10keV) observations. The dashed, dotted and solid lines are
narrow, wide jet and combined calculation.} \label{Fig:fig2}
\end{figure}

\end{document}